# A practical guide to the simultaneous determination of protein structure and dynamics using metainference

Thomas Löhr[1], Carlo Camilloni[2], Massimiliano Bonomi[3], Michele Vendruscolo[1,*]

[1]Department of Chemistry, University of Cambridge, CB2 1EW Cambridge, UK

[2]Dipartimento di Bioscienze, Università degli Studi di Milano, 20133 Milano, Italy

[3]Structural Bioinformatics Unit, Institut Pasteur, CNRS UMR 3528, 75015 Paris, France

[*]Corresponding author: mv245@cam.ac.uk

**Summary**

Accurate protein structural ensembles can be determined with metainference, a Bayesian inference method that integrates experimental information with prior knowledge of the system and deals with all sources of uncertainty and errors as well as with system heterogeneity. Furthermore, metainference can be implemented using the metadynamics approach, which enables the computational study of complex biological systems requiring extensive conformational sampling. In this chapter, we provide a step-by-step guide to perform and analyse metadynamic metainference simulations using the ISDB module of the open-source PLUMED library, as well as a series of practical tips to avoid common mistakes. Specifically, we will guide the reader in the process of learning how to model the structural ensemble of a small disordered peptide by combining state-of-the-art molecular mechanics force fields with nuclear magnetic resonance data, including chemical shifts, scalar couplings and residual dipolar couplings.

**Key Words**

functional dynamics; ensemble determination; Bayesian data modelling; integrative modelling;

**Running Head Title:** A practical guide to metainference

# 1. Introduction

The goal of molecular dynamics (MD) simulations is to provide a characterization of molecular processes in terms of their structures, thermodynamics and kinetics *(1)*. This goal is ambitious and presents a series of major challenges, including the development of accurate force fields, of effective sampling methods and of quantitative accounting of the various sources of errors *(2-4)*. Quite generally, because of the intrinsic approximations of the force fields and the fact that they are optimized to maximize transferability (see Chapters 1 to 3), the results of a simulation of a specific system might not quantitatively match available experimental measurements, even when sampling is exhaustive *(5, 6)*. In Parts I and II of this book, the reader is introduced to the strengths and weaknesses of current force fields, as well as to some of the approaches for achieving exhaustive sampling of the conformational space.

Nonetheless, the advances made over a period of 50 years are making it possible to provide atomistic interpretations of various experimental measurements of molecular processes in terms of structural ensembles. For example, one might want to determine the distribution of configurations underlying a small-angle X-ray scattering (SAXS) profile or a nuclear magnetic resonance (NMR) chemical shifts spectrum of a protein. In order for this interpretation to be significant, the structural ensemble must be well defined in terms of statistical mechanics, as well as in quantitative agreement with the available experimental observations *(2, 3, 7, 8)*.

Alternatively, given a set of equilibrium measurements, one can ask whether there is a representative structure, or more generally and ensemble of structures, that explains them. This is a typical inverse problem that can be solved by using some form of regularisation technique *(8, 9)*. An accurate force field combined with exhaustive sampling will provide an ensemble of structures that quantitatively match the experimental data, or, if appropriate, a representative structure that summarises the main structural features of the system. However, as mentioned above, such an ideal force field does not yet exist, and therefore one can ask: given a state-of-the-art transferable force field and a specific system of interest, how can we obtain a quantitative agreement with the available experimental knowledge? In this section and starting from the present chapter, we will explore different approaches to address this problem.

## 2. Theory

In MD force fields, when one wants to enforce a property $f(X)$ on a given conformation $X$ (e.g. the length of a covalent bond), a harmonic potential, e.g. $k \cdot (f(X) - d)^2$, can be used to restrain the property around the expected value $d$ with strength $k$. However, many properties measured using bulk techniques cannot be expressed in terms of individual conformations, but they can only be calculated on the ensemble of conformations populated under certain external conditions. For example, solution measurements, such as SAXS and many NMR observables, depend on the average of specific structural properties calculated over the entire ensemble. In these situations, one can approximate the ensemble with a certain number of copies of the system (replicas), calculate the property $f(X)$ for each replica, and then apply the restraint to the average $\langle f(X) \rangle$ across the replicas as $k \cdot (\langle f(X) \rangle - d)^2$ *(10, 11)*.

To complicate this picture, experimental measurements of any property of the system are affected by random noise and systematic errors *(7, 12)*. Furthermore, also the structural interpretation of an experimental observable (the function $f(X)$, also known as *forward model* or *predictor*) is hampered by approximations that might introduce additional errors. As a consequence, the agreement between experimental measurements and the ensemble obtained from the simulation should be enforced only up to a certain extent *(2, 3, 7, 8)*, which can be quantified by the overall error. In practical situations, estimating this error is far from trivial.

Statistical inference offers a rigorous theoretical framework to combine all the available sources of information about a system in order to obtain an accurate and precise description of its properties. The metainference approach *(2)* described in this chapter, by building on the inferential structure determination method *(13)*, enables modelling accurate structural ensembles by optimally combining prior information on a system with noisy, ensemble-averaged experimental data and by keeping into account all sources of errors introduced above *(8)*.

## 2.1 Inferential Structure Determination

To construct the best possible structural model of a system, one can score different possible models according to their consistency with the overall knowledge available. This includes theoretical knowledge (often called the 'prior' information, *I*), such as physico-chemical information about the system (the force field), and the knowledge acquired from experimental measurements (i.e. the 'data', $D$). In this view, the best model is the one that is most likely to occur given the information available.

Inferential structure determination (ISD) *(13)* is a Bayesian inference approach that, by estimating the probability of a model given the information available, enables one to infer the best possible model. In this approach, the assessment of the quality of a model, $M$, is made with the *posterior probability* $p(M|D,I)$ of $M$ given $D$ and $I$, which is given by

$$p(M|D,I) \propto p(D|M,I)\, p(M|I) \quad (1)$$

where the *likelihood function* $p(D|M,I)$ is the probability of observing $D$ given $M$ and $I$, and the *prior probability* $p(M|I)$ is the probability of $M$ given $I$.

To define the likelihood function, one needs a *forward model* $f_i(X)$ to predict the data point $d_i$ that would be observed for a system in state $X$, and a *noise model* that specifies the distribution of the deviations between observed and predicted data. Both the forward model and the noise model are defined in terms of unknown parameters that are part of the model $M$ and inferred along with the state $X$ by sampling the posterior distribution. The sampling is usually carried out using Monte Carlo (MC), MD, or combined techniques based on Gibbs sampling *(14)*.

ISD has been used to determine the macromolecular architecture of several protein complexes of outstanding biological importance *(15-23)*, using a variety of experimental data and software such as the Integrative Modelling Platform (IMP) *(24)* and the Crystallography & NMR System (CNS) *(25)*.

**2.2 Metainference**

Metainference *(2)* extends ISD *(13)* to deal with experimental data that are averaged over multiple conformations and thus enables modelling structural ensembles *(8)*. In metainference, the modified force field (or metainference energy function) for a set of $N$ replicas of the system, is defined as $E_{MI} = -k_B T \log p_{MI}$, where $k_B$ is the Boltzmann constant, $T$ the temperature of the system, and $p_{MI}$ the metainference posterior probability. In general terms, the metainference energy function can be written as

$$E_{MI} = -k_B T \log p_{MI}(\boldsymbol{X}, \tilde{\boldsymbol{f}}, \boldsymbol{\sigma}^{SEM}, \boldsymbol{\sigma}^B | \mathbf{D})$$

$$= -k_B T \log \left[ \prod_{r=1}^{N} p(X_r) \prod_{i=1}^{N_d} p(d_i | \tilde{f}_{r,i}, \sigma_{r,i}^B) \, p(\tilde{f}_{r,i} | \mathbf{X}, \sigma_{r,i}^{SEM}) \, p(\sigma_{r,i}^{SEM}) \, p(\sigma_{r,i}^B) \right] \quad (2)$$

where:

- $\mathbf{D} = [d_i]$ is a set of $N_d$ independent experimental data points;
- $\boldsymbol{X} = [X_r]$, where $X_r$ represents the state of replica $r$, defined here by the coordinates of all the particles of the system;

- $\tilde{\boldsymbol{f}} = [\tilde{f}_{r,i}]$, where $\tilde{f}_{r,i}$ is the average of the predictor (forward model) $f_i$ of the $i$-th experimental observable, calculated over an infinite number of replicas;

- $\boldsymbol{\sigma}^{SEM} = [\sigma_{r,i}^{SEM}]$, where $\sigma_{r,i}^{SEM}$ is the standard error of the mean related to the average of $f_i$ being calculated over a finite number of replicas;

- $\boldsymbol{\sigma}^B = [\sigma_{r,i}^B]$, where $\sigma_{r,i}^B$ is an uncertainty parameter that describes random and systematic errors in the experimental data point $d_i$ as well as in the forward model $f_i$;

- $p(d_i|\tilde{f}_{r,i}, \sigma_{r,i}^B)$ encodes the noise model (data likelihood), defined as the conditional probability of $d_i$ given $\tilde{f}_{r,i}$ and $\sigma_{r,i}^B$;

- $p(\tilde{f}_{r,i}|\mathbf{X}, \sigma_{r,i}^{SEM})$ is the conditional probability of observing $\tilde{f}_{r,i}$ given that the average of $f_i$ is calculated on a finite number of replicas $N$, $f_i(\mathbf{X}) = \frac{1}{N}\sum_{r=1}^N f_i(X_r)$. According to the central limit theorem (CLT), this is a Gaussian distribution;

- $p(\sigma_{r,i}^{SEM})$ encodes the CLT scaling of $\sigma_{r,i}^{SEM}$ with $N$: $\sigma_{r,i}^{SEM} \propto 1/\sqrt{N}$.

- $p(\sigma_{r,i}^B)$ and $p(X_r)$ are the priors on $\sigma_{r,i}^B$ and $X_r$, respectively.

For simplicity, in the following we will consider the specific case of Gaussian noise. However, all the considerations below remain valid in the general case of Eq. 2. When the data likelihood $p(d_i|\tilde{f}_{r,i}, \sigma_{r,i}^B)$ is a Gaussian function, the metainference energy function $E_{MI}$ can be written as *(2)*

$$E_{MI} = E_{FF} + \frac{k_B T}{2} \sum_{r,i} \frac{[d_i - f_i(\boldsymbol{X})]^2}{\left(\sigma_{r,i}^B\right)^2 + \left(\sigma_{r,i}^{SEM}\right)^2} + E_\sigma \quad (3)$$

where the force field of standard MD simulations $E_{FF} = \sum_{r=1}^{N} E_{FF}(X_r) = -k_B T \sum_{r=1}^{N} \log p(X_r)$ is modified by *i)* a series of (harmonic) *data-restraints*, which enforce the agreement of the replicas with the ensemble-averaged data, and *ii)* an *error restraint*, $E_\sigma = k_B T \sum_{r,i} \left\{ -\log p(\sigma_{r,i}^B) + 0.5 \log \left[ \left(\sigma_{r,i}^B\right)^2 + \left(\sigma_{r,i}^{SEM}\right)^2 \right] \right\}$, that accounts for normalization of the data likelihood and error priors.

Metainference has been used to model structural ensembles using multiple NMR data *(26, 27)* and more recently cryo-electron microscopy density maps *(28, 29)*. Furthermore, the metainference equivalence to ISD has been used to perform an integrative structure refinement of a protein RNA-complex using SAXS and NMR data *(30)*.

## 2.3 Implementation

### 2.3.1 Gibbs sampling

In the following, we describe how a metainference simulation is run in practice. Given the system of interest, multiple MD simulations (the replicas) are prepared using the same force field and simulation setup (number of atoms, temperature, pressure, etc.). The replicas are then simulated in parallel using the energy function in Eq. 3. At each time step, the metainference energy is calculated as the sum of the force-field energy of all the replicas, the data-restraints, and the error-restraints. The intensity of the data-restraint is ultimately determined by the value of the errors parameters $\sigma_{r,i}^{B}$, which quantify the level of noise: small errors will result in strong structural restraints; outliers and high-error data points will automatically decrease the strength of the data-restraint. The conformations $X$ and the error parameters are then updated using a Gibbs sampler, as described in Fig. 1.

### 2.3.2 Parameters optimization

In a metainference simulation, the number of replicas employed is necessarily smaller than the numbers of conformations actually contributing to an experimental observable. This discrepancy is accounted for by the variable $\sigma_{r,i}^{SEM}$ in Eq. 3, which quantifies the error in calculating averaged properties using a small set of replicas. According to the central limit theorem, $\sigma_{r,i}^{SEM}$ is proportional to $1/\sqrt{N}$. This term can be estimated on-the-fly *(31)* as the standard error of the mean over the replicas, which can be calculated either on the entire trajectory or on a window of a finite size

$$\sigma_i^{SEM} = \sqrt{\sum_{r=1}^{N} \frac{(f_{r,i}(X) - \langle f_{r,i}(X) \rangle)^2}{N}} \quad (4)$$

**2.4 Metadynamic metainference**

As in standard MD simulations, in metainference simulations relevant states might be separated by large free-energy barriers. To accelerate sampling, metainference was combined with metadynamics *(32)* in its Parallel-Bias (PBMetaD) flavour *(33)* (see Note 1). In this combined approach (M&M) *(34)*, an additional, time-dependent bias potential $V_{PB}$ is added to each replica and shared among all of them, in the spirit of the multiple-walkers approach *(35)*. As a consequence, one need to account for the $V_{PB}$ bias potential when calculating the average forward model $f_i(X)$ used in the metainference data-restraint. A weighted average can be calculated using the Umbrella Sampling reweighting weights *(36)*, which instantaneously accounts for the presence of the PBMetaD bias potential (see Note 2). Furthermore, these weights can be averaged over a short time window, in order to decrease their fluctuations and prevent numerical instabilities due to too high instantaneous forces. As a consequence of using a weighted average to calculate $f_i(X)$, the effective number of replicas might vary during the simulation along with the associated error $\sigma_{r,i}^{SEM}$. To account for this effect, $\sigma_{r,i}^{SEM}$ can be estimated as the standard error of the weighted mean *(31)*.

## 3. Materials

Simulations of the EGAAWAASS peptide were carried out using GROMACS 5.1.4 *(37)* and the ISDB module *(38)* of the PLUMED open-source library, version 2.3 *(39)*. For didactic purposes, the scripts presented here are updated to PLUMED version 2.5. The initial conformation of the peptide was modeled using VMD *(40)* and all plots were created with the Matplotlib library *(41)*. All simulations should be run in parallel on a cluster machine using MPI. The reader should refer to the GROMACS and PLUMED user manuals for detailed instructions about how to compile and execute the codes. Basic knowledge about the use of GROMACS is required to setup the MD simulations and manipulate the trajectories.

## 4. Methods

In this section, we will demonstrate the use of M&M *(34)* on the EGAAWAASS peptide *(31)*. This molecule is highly disordered and has been used as a model system to study the quality of MD force fields *(42)* and the suitability of residual dipolar couplings (RDCs) to reveal structural information *(43)*. The quality of modern force fields is insufficient to accurately determine the structural ensemble of this system, making this an excellent candidate for the application of M&M. Previous NMR studies *(43)* provided chemical shifts, 3J-couplings and RDCs data (Tables 1, 2, 3), which can be used with M&M to correct the inaccuracies of the underlying force field. By comparing simulations performed with increasing amounts of experimental data, we can evaluate the impact of specific experimental observables on the accuracy of the reconstructed ensemble.

This section is organized as follows. In subsection 4.1, we describe the system preparation and equilibration steps. In subsection 4.2, we simulate the system using PBMetaD without the addition of experimental data. In subsection 4.3, we introduce chemical shifts and 3J-couplings as experimental restraints in the M&M framework. We describe the setup of the simulation as well as the various parameters and observables that need to be monitored during the simulation. In subsection 4.4, we also add RDCs, which require some additional considerations. Finally, in subsection 4.5, we discuss the protocol used to analyze the simulations, such as the calculation of root-mean-square deviations (RMSDs) from experimental data, and the free energy surfaces generated by all simulations.

4.1 System preparation

The EGAAWAASS peptide is initially modelled with VMD and solvated in a rhombic dodecahedron box with side lengths of 4.5, 4.5, and 3.2 nm and containing 2118 water molecules. The system is neutralized by 3 $Na^+$ and 2 $Cl^-$ ions. Energy minimization of the system is performed using the steepest descent algorithm to a maximum force of less than 100 kJ/(mol/nm). Equilibration is performed for 500 ps in the NVT ensemble using the Bussi-Donadio-Parrinello thermostat *(44)* and for 500 ps in the NPT ensemble using the Parrinello-Rahman barostat *(45),* with position restraints added to all heavy atoms. We use the CHARMM22* force field *(46)* in combination with the TIP3P water model. We also use the Particle-Mesh-Ewald *(47)* approach for both van der Waals and electrostatic interactions with a cut-off of 0.9 nm, as well as the LINCS algorithm *(48)* for constraint solving using a matrix expansion on the order of 6 and 2 iterations per step (see Note 3).

4.2 PBMetaD simulation

We begin with simulating the EGAAWAASS peptide without the addition of experimental data. To ensure an adequate sampling of the conformational landscape of this system, we use well-tempered *(49)* PBMetaD *(33)* (see Note 1), in combination with the multiple-walkers approach *(35)*. We will use all the backbone dihedral angles $\phi$ and $\psi$ as CVs, as well as the W5 $\chi^1$ and $\chi^2$ dihedral angles, the similarities (`DIHCOR`) of the dihedral angles between each pair of alanine residues and the E1-S9 $C^\alpha$-$C^\alpha$ distance. The following PLUMED input file can be used to define the CVs listed above:

```
MOLINFO MOLTYPE=protein STRUCTURE=egaawaass.pdb
WHOLEMOLECULES ENTITY0=1-111

# Dihedral backbone angles: Psi9, Phi1 are not defined
psi1: TORSION ATOMS=@psi-1
psi2: TORSION ATOMS=@psi-2
psi3: TORSION ATOMS=@psi-3
psi4: TORSION ATOMS=@psi-4
psi5: TORSION ATOMS=@psi-5
psi6: TORSION ATOMS=@psi-6
psi7: TORSION ATOMS=@psi-7
psi8: TORSION ATOMS=@psi-8
phi2: TORSION ATOMS=@phi-2
phi3: TORSION ATOMS=@phi-3
phi4: TORSION ATOMS=@phi-4
phi5: TORSION ATOMS=@phi-5
phi6: TORSION ATOMS=@phi-6
phi7: TORSION ATOMS=@phi-7
phi8: TORSION ATOMS=@phi-8
phi9: TORSION ATOMS=@phi-9

# Bulky Trp residue dihedral
dihtrp_cacb: TORSION ATOMS=67,47,49,52
dihtrp_cbcg: TORSION ATOMS=47,49,52,53

# Similarity of Ala-Ala dihedrals
aasimpsi: DIHCOR ATOMS1=@psi-3,@psi-6
aasimphi: DIHCOR ATOMS1=@phi-4,@phi-7

# Distance between alpha-carbons of first and last residue
peplen: DISTANCE NOPBC ATOMS=5,102
```

To use the @ shorthand to define the four atoms of the TORSION CV, we need to first specify a structure file with the MOLINFO directive. A convenient way is to generate a PDB file from the standard GROMACS TPR file:

```
gmx make_ndx -f topol0.tpr
gmx editconf -f topol0.tpr -n index.ndx -o egaawaass.pdb
```

The first command creates an index file, which will allow us to select only the protein atoms in the second line. The WHOLEMOLECULES command tells PLUMED to rebuild molecules that have been broken inside the MD code by periodic boundary conditions (see Note 4). We can now setup the PBMETAD directive using the CVs previously defined as arguments (`ARG`). We will choose a `BIASFACTOR` of 8, a Gaussian deposition `PACE` of 1 ps, an initial Gaussian `HEIGHT` of 0.3 kJ/mol, and Gaussian widths `SIGMA` equal to 0.6 rad for the dihedrals, 0.3 for the dihedral similarities and 0.3 nm for the end-to-end distance. We use the `WALKERS_MPI` flag to instruct PLUMED to activate the multiple-walkers approach and share the bias across replicas using MPI.

```
PBMETAD ...
ARG=phi2,phi3,phi4,phi5,phi6,phi7,phi8,phi9,psi1,psi2,psi3,psi4,psi5,psi6,psi7,psi8,dihtrp_cacb,dihtrp_cbcg,aasimpsi,aasimphi,peplen
SIGMA=0.6,0.6,0.6,0.6,0.6,0.6,0.6,0.6,0.6,0.6,0.6,0.6,0.6,0.6,0.6,0.6,0.6,0.3,0.3,0.3
HEIGHT=0.3
PACE=500
BIASFACTOR=8
LABEL=pb
GRID_MIN=-pi,-pi,-pi,-pi,-pi,-pi,-pi,-pi,-pi,-pi,-pi,-pi,-pi,-pi,-pi,-pi,-pi,-pi,0,0,0
GRID_MAX=pi,pi,pi,pi,pi,pi,pi,pi,pi,pi,pi,pi,pi,pi,pi,pi,pi,pi,1,1,3.5
WALKERS_MPI
... PBMETAD
```

The grid options (`GRID_MIN` and `GRID_MAX`) allow us to store the bias on a grid, thus increasing the computational performances. The value of the PBMetaD bias potential and the associated forces in a generic point of the CV space are then calculated using a bicubic spline interpolation of the grid points. The units of measures are: kJ/mol for energy, nm for distances, K for temperature, and number of MD steps for time (here the time step is set to 2 fs). Finally, we print out the value of each biased CV as well as the PBMetaD bias.

```
PRINT
ARG=phi2,phi3,phi4,phi5,phi6,phi7,phi8,phi9,psi1,psi2,psi3,psi4,psi5,psi6,psi7,psi8,dihtrp_cacb,dihtrp_cbcg,aasimpsi,aasimphi,peplen,pb.bias FILE=CVS STRIDE=500
```

We are now ready to start the simulation. Starting from 14 different conformations

extracted from the equilibration run, we generate 14 TPR files and run the following command (see Note 5):

```
mpirun -n 14 gmx_mpi mdrun -s topol -plumed plumed.dat -multi 14
```

We let the simulation run until convergence (see Note 6) and then perform a more thorough analysis (Section 4.5).

4.3 M&M with 3J-couplings and chemical shifts

We now simulate the EGAAWAASS peptide using 3J-couplings and chemical shifts. In order to do this, we need to add to the PLUMED file described above the forward models of the experimental data that will be incorporated into the M&M simulation. 3J-couplings are related to the backbone dihedral angles through the Karplus equation *(50)*

$$^3J(\theta) = A\cos^2(\theta + \Delta\theta) + B\cos(\theta + \Delta\theta) + C \quad (5)$$

where $\theta$ is the dihedral angle in question (either $\phi$ or $\psi$), while A, B, C and $\Delta\theta$ are empirically-determined parameters, which depend on the type of coupling observed. PLUMED allows us to calculate these experimental observables by using the `JCOUPLING` directive, and to specify the reference (experimental) values by adding the `ADDCOUPLINGS` flag. We also need to specify the `TYPE` of coupling *(51, 52)* and list the dihedral angles associated to each coupling. In the following PLUMED input, we define the H$\alpha$-N, H$\alpha$-HN, C-C$\gamma$ and N-C$\gamma$ 3J-couplings along with their reference values in Hz.

```
# _G_AW_AS_
JCOUPLING ...
    ADDCOUPLINGS
    TYPE=HAN
    ATOMS1=@psi-2 COUPLING1=-0.49
    ATOMS2=@psi-4 COUPLING2=-0.54
    ATOMS3=@psi-5 COUPLING3=-0.53
    ATOMS4=@psi-7 COUPLING4=-0.39
    ATOMS5=@psi-8 COUPLING5=-0.39
    LABEL=jhan
... JCOUPLING

# __AAWAASS
JCOUPLING ...
    ADDCOUPLINGS
    TYPE=HAHN
    ATOMS1=@phi-2 COUPLING1=6.05
    ATOMS2=@phi-3 COUPLING2=5.95
    ATOMS3=@phi-4 COUPLING3=6.44
    ATOMS4=@phi-5 COUPLING4=6.53
    ATOMS5=@phi-6 COUPLING5=5.93
    ATOMS6=@phi-7 COUPLING6=6.98
    ATOMS7=@phi-8 COUPLING7=7.16
    LABEL=jhahn
... JCOUPLING

# ____W____
JCOUPLING ...
    ADDCOUPLINGS
    TYPE=CCG
    ATOMS1=@chi1-5 COUPLING1=1.59
    LABEL=jccg
... JCOUPLING

# ____W____
JCOUPLING ...
    ADDCOUPLINGS
    TYPE=NCG
    ATOMS1=@chi1-5 COUPLING1=1.21
    LABEL=jncg
... JCOUPLING
```

We now add the chemical shifts, which are implemented in the PLUMED action activated by the `CS2BACKBONE` directive, This action uses the CamShift algorithm *(53)* to calculate the chemical shifts from a given structure using the following equation

$$\delta_a^{\text{pred}} = \delta_a^{\text{rc}} + \sum_{b,c} \alpha_{bc} d_{bc}^{\beta_{bc}} \quad (6)$$

where $\delta_a^{\text{pred}}$ is the predicted chemical shift of atom $a$, $\delta_a^{\text{rc}}$ is the random coil chemical shift of atom $a$, $d_{bc}$ is the distance between atoms $b$ and $c$ and $\alpha_{bc}$ and $\beta_{bc}$ are atom- and residue-dependent empirical parameters. Atoms $b$ and $c$ are chosen based on complex distance and residue criteria *(53)*. In the line of the CS2BACKBONE directive, we need to specify the ATOMS involved in the calculation of the chemical shifts, which are typically all the atoms of the protein:

```
cs: CS2BACKBONE ATOMS=1-111 DATADIR=data TEMPLATE=egaawaass.pdb
```

For computational efficiency, PLUMED internally uses a neighbour list to calculate the pairwise interactions required by CS2BACKBONE. Furthermore, we need to supply a PDB file of the molecule (referred to as TEMPLATE), as well as the name of the data folder (DATADIR). This folder contains the reference chemical shifts (#shifts.dat), the reference structure (egaawaass.pdb) and the CamShift

```
data
├── CAshifts.dat
├── CBshifts.dat
├── Cshifts.dat
├── HAshifts.dat
├── Hshifts.dat
├── Nshifts.dat
├── camshift.db
└── egaawaass.pdb
```

database (camshift.pdb):

We are now ready to setup the metainference calculations using the following input:

```
METAINFERENCE ...

ARG=(cs\.nh_.*),(cs\.hn_.*),(cs\.ha_.*),(cs\.ca_.*),(cs\.cb_.*),
(cs\.co_.*),(jhan\.j_.*),(jhahn\.j_.*),(jccg\.j.*),(jncg\.j.*),p
b.bias

PARARG=(cs\.expnh.*),(cs\.exphn.*),(cs\.expha.*),(cs\.expca.*),(
cs\.expcb.*),(cs\.expco.*),(jhan\.exp_.*),(jhahn\.exp_.*),(jccg\
.exp.*),(jncg\.exp.*)
    NOISETYPE=MGAUSS
    REWEIGHT
    OPTSIGMAMEAN=SEM AVERAGING=200
    SIGMA0=25.0 SIGMA_MIN=0.001 SIGMA_MAX=25.0 DSIGMA=0.1
    WRITE_STRIDE=10000
    LABEL=bycsj
... METAINFERENCE
```

We will go through the `METAINFERENCE` directive line by line. First, in `ARG` we specify the output of our previously defined CVs representing the experimental observables. `CS2BACKBONE` sorts the different types of chemical shifts into different components and residue numbers, so we can use regular expressions to conveniently provide a list of these CVs. The same principle applies to the 3J-couplings. A very important part of this line is the addition of the PBMetaD bias from the `PBMetaD` directive (`pb.bias`) at the end of the list of arguments. This bias is used to calculate a weighted ensemble average of the experimental observables by accounting for the PBMetaD bias potential (see Section 2.4). Then, in `PARARG` we specify the experimental reference values, in the same order as in the `ARG` keyword and again using regular expressions.

We continue by specifying the `NOISETYPE`. We assume that data points are independent and we use a Gaussian model of noise (`NOISETYPE`) with one error parameter per data point (`MGAUSS`). Other available options are a single error parameter for all data points (GAUSS) or long-tailed distributions to account for outliers (OUTLIERS, MOUTLIERS). The latter can be used when large random or systematic errors are expected for a few data points. The REWEIGHT flag indicates that we are passing to the METAINFERENCE directive an additional argument (the last) in ARG, which contains the value of the PBMetaD bias. Next, we specify the technique used for calculating the standard error of the mean $\sigma_{r,i}^{SEM}$ (OPTSIGMAMEAN). In this case SEM implies the automatic estimation via a windowed average calculation, in which the size of the window in steps is given by AVERAGING. This quantity corresponds also the size of the window used to average the weights from PBMetaD.

We sample the error associated with each data using a MC algorithm (see Note 7). We specify a starting value SIGMA0, lower and upper sampling bounds SIGMA_MIN and SIGMA_MAX and a step size DSIGMA. As the data-restraint force is inversely proportional to both the standard error of the mean $\sigma_{r,i}^{SEM}$ and the Bayesian error $\sigma_{r,i}^{B}$ and that the averaging procedure for the estimation of $\sigma_{r,i}^{SEM}$ may start from very low values (if the starting configurations are similar), it is thus safer to begin the sampling of $\sigma_{r,i}^{B}$ from a fairly high value. The sampling range of $\sigma_{r,i}^{B}$ varies depending on the type of experimental data used (see Note 8). Finally, we allow METAINFERENCE to create checkpoint files every 10000 steps with WRITE_STRIDE. These files contain information necessary to restart the simulations, such as the variances of each experimental observable, as well as the last values of the errors $\sigma_{r,i}^{B}$, so that PLUMED can restart sampling from where it left off.

Before running the simulation, we can instruct PLUMED to calculate some relevant information that are useful to monitor the simulation:

```
# Calculate weighted ensemble average
ENSEMBLE ...

ARG=(nh\.rdc_.*),(caha\.rdc_.*),(jhan\.j_.*),(jhahn\.j_.*),(jccg
\.j_.*),(jncg\.j_.*),(cs\..._.*),pb.bias REWEIGHT
    LABEL=ens
... ENSEMBLE

# We use the analogous function for all other observables
STATS ...
    ARG=(ens\.cs\..._.*) PARARG=(cs\.exp.*)
    LABEL=csst
... STATS
```

`ENSEMBLE` performs the same action as the first line in `METAINFERENCE`, i.e. it calculates a weighted ensemble average on-the-fly. The `STATS` directive calculates useful statistical information, such as the correlation with the experimental values, thus allowing us to quickly judge the quality of our ensemble. Finally, we print out the output of the `STATS` and `ENSEMBLE` directives above and the value of each CV biased by PBMetaD:

```
PRINT ARG=bycsj.*                  STRIDE=100 FILE=BAYES.CSJ
PRINT ARG=csst.*,(ens\.cs\..._.*) STRIDE=500 FILE=ST.CS
PRINT
ARG=phi2,phi3,phi4,phi5,phi6,phi7,phi8,phi9,psi1,psi2,psi3,psi4,
psi5,psi6,psi7,psi8,dihtrp_cacb,dihtrp_cbcg,aasimpsi,aasimphi,pe
plen,pb.bias FILE=CVS STRIDE=500
```

The output from metainference contains the values of all the errors ($\sigma_{r,i}^{B}$ and $\sigma_{r,i}^{SEM}$), information regarding the sampling of these quantities, the weight of each replica and the metainference energy. We start the simulation as we previously did (Section 4.2).

M&M makes use of multiple replicas of the system and, generally speaking, a higher number of replicas leads to a higher quality result (see Note 9). To monitor the simulation on-the-fly and ensure the effectiveness of the metainference approach, we can look at the value of the standard error of the mean $\sigma_{r,i}^{SEM}$ (Fig. 2) and of the error $\sigma_{r,i}^{B}$ (Fig. 3) along the simulation. These two quantities together determine the overall intensity of the data-restraint. Furthermore, we can monitor the metainference energy, its derivative with respect to the PBMetaD bias, the MC acceptance rate of the error parameters $\sigma_{r,i}^{B}$ (Fig. 4), and the agreement with the experimental data during the simulation (Fig. 5, see Note 10). We will let the simulation run until convergence (see Note 6) and then perform a more thorough analysis (Section 4.4).

## 4.4 M&M simulation 3J-couplings, chemical shifts, and RDCs

We continue by adding residual dipolar couplings (RDCs) to the M&M simulation. RDCs can be calculated using the $\theta$-method *(54)*:

$$D_i = -\frac{\mu_0 \gamma_1 \gamma_2 \hbar}{8\pi^3} \langle \frac{3\cos^2 \vartheta_i - 1}{r_i^3} \rangle \quad (7)$$

Here, $D_i$ is the residual dipolar coupling, $r_i$ is the bond length, $\vartheta_i$ is the angle between the bond in question and the external magnetic field (usually taken to be the z-axis), $\mu_0$, $\gamma_1$ and $\gamma_2$ are atom dependent constants, and $\hbar$ is the Planck constant. RDCs are measured in alignment media and report on the fraction of aligned molecules. Thus, directly comparing these experimentally observed values with those calculated in a simulation makes little sense. The relationship is governed by a scaling factor $\lambda$ that can be sampled during the simulation.

We start by defining our forward models for C$\alpha$-H$\alpha$ and N-H RDCs by using the `RDC` PLUMED directive. As in the case of the 3J-couplings, we use the `ADDCOUPLINGS` flag to enable adding the reference experimental values. For each measured RDC, we need to specify a pair of `ATOMS` and the corresponding experimental value (`COUPLING`). We also specify the gyromagnetic ratio $\gamma$ with the `GYROM` keyword, which is dependent on the type of observed dipolar coupling (see PLUMED manual). Finally, the `SCALE` keyword allows us to apply a fixed rescaling factor to the calculated RDCs.

```
# _GAAWAASS
RDC ...
    ADDCOUPLINGS
    GYROM=-72.5388
    SCALE=0.0001
    ATOMS1=18,19 COUPLING1=-5.4
    ATOMS2=25,26 COUPLING2=-1.26
    ATOMS3=35,36 COUPLING3=-5.22
    ATOMS4=45,46 COUPLING4=-0.91
    ATOMS5=69,70 COUPLING5=2.33
    ATOMS6=79,80 COUPLING6=-2.88
    ATOMS7=89,90 COUPLING7=-8.37
    ATOMS8=100,101 COUPLING8=-3.78
    LABEL=nh
... RDC

# E_AAWAAS_
RDC ...
    ADDCOUPLINGS
    GYROM=179.9319
    SCALE=0.0001
    ATOMS1=5,6 COUPLING1=12.95
    ATOMS2=27,28 COUPLING2=11.5
    ATOMS3=37,38 COUPLING3=21.42
    ATOMS4=47,48 COUPLING4=-9.37
    ATOMS5=71,72 COUPLING5=10.01
    ATOMS6=81,82 COUPLING6=15.01
    ATOMS7=91,92 COUPLING7=15.73
    LABEL=caha
... RDC
```

We are now ready to define the METAINFERENCE directive. We will use a similar setup as the one used above for the chemical shifts, with the addition of the keywords needed to activate the sampling of the scaling factor. As we are using two different datasets of RDCs, we need to use two separate METAINFERENCE directives to allow for two different scaling factors. In each directive, we use the SCALEDATA flag to indicate the use of a variable scaling factor, and the SCALE_PRIOR flag to define the type of prior distribution. As the scaling factor should remain relatively constant over time, specifying a Gaussian prior (see Note 11) will allow us to sample points around a typical value `SCALE0` without deviating too much from this estimate. The standard deviation of this Gaussian prior is specified with `DSCALE`.

```
METAINFERENCE ...
    ARG=(nh\.rdc_.*),pb.bias
    PARARG=(nh\.exp_.*)
    NOISETYPE=MGAUSS
    SCALEDATA
    REWEIGHT
    OPTSIGMAMEAN=SEM AVERAGING=200
    SCALE_PRIOR=GAUSSIAN SCALE0=8.0 DSCALE=0.5
    SIGMA0=25.0 SIGMA_MIN=0.001 SIGMA_MAX=25.0 DSIGMA=0.1
    WRITE_STRIDE=10000
    LABEL=byrdcnh
... METAINFERENCE

METAINFERENCE ...
    ARG=(caha\.rdc_.*),pb.bias
    PARARG=(caha\.exp_.*)
    NOISETYPE=MGAUSS
    SCALEDATA
    REWEIGHT
    OPTSIGMAMEAN=SEM AVERAGING=200
    SCALE_PRIOR=GAUSSIAN SCALE0=9.0 DSCALE=0.5
    SIGMA0=25.0 SIGMA_MIN=0.001 SIGMA_MAX=25.0 DSIGMA=0.1
    WRITE_STRIDE=10000
    LABEL=byrdccaha
... METAINFERENCE
```

Estimating the correct value for the scaling factor can be done as follows. First, one should set `DSCALE` to some fairly large number, and set `SCALE0` to some arbitrary starting point. By running a short simulation (in this case 100 ps) and monitoring the value of the scaling factor in the output of the `METAINFERENCE` directive, we can obtain a measure of a reasonable sampling range. Then, the static scaling in the RDC CV can be adjusted together with the starting point `SCALE0` and `DSCALE`. This whole procedure should be done separately for each individual RDC dataset, and the resulting factors verified when both datasets are active, as the scaling factors can be subtly influenced by additional restraints. Once the scaling factor $\lambda$ has been correctly determined, we should expect to see values for $\lambda$ oscillating around our `SCALE0` value (Fig. 6A), along with a fairly high MC acceptance rate (Fig. 6B). If our initial guess is inaccurate, we will see a low acceptance rate together with sampled values that tend to be far away from our initial guess. In this case, we would have to revise our estimate of the scaling factor (see Note 12).

We run the production simulation in the same way as before. We will monitor the errors $\sigma_{r,i}^{B}$ and $\sigma_{r,i}^{SEM}$ as well as the other metainference observables and the correlations between the forward models and the reference experimental data. We should also monitor the value of the scaling factor $\lambda$ (Fig. 6) and, if necessary, make any adjustments to the sampling range.

4.5 Analysis

In the analysis of our simulations, we will focus on the experimental observables and their associated errors and also briefly illustrate how to calculate probability distributions for any generic CVs. First, we should concatenate our trajectories:

```
$ gmx trjcat -f traj_comp* -o cat_traj.xtc -settime
$ gmx trjconv -f cat_traj.xtc -s topol0.tpr -o traj.xtc -pbc mol
```

The `settime` flag allows us to specify the starting and end time for each replica's trajectory. It should be used to obtain one continuous trajectory. We also correct discontinuities due to periodic boundary conditions and remove the water, if present. To analyse this resulting trajectory, we will make use of the PLUMED `driver` utility, which reads in a trajectory and calculates certain observables based on those frames. To do this, the driver requires a PLUMED input file, very similar to the one used in the simulation, with some important differences. First, we need to pass the `RESTART` flag to PLUMED. Secondly, we need to adjust the `PACE` parameter in PBMetaD to stop PLUMED from adding additional Gaussians to the `HILLS` files and pass the simulation temperature to PLUMED by adding `TEMP=300` (see Note 13). We are especially interested in the PBMetaD bias per frame, as we need it to calculate the weights *(36)*. We will also calculate the radius of gyration $R_g$ of the peptide for each frame, by using the `GYRATION` directive and specifying all Cα carbons as arguments. Finally, we calculate the experimental observables for each frame using the same directives defined in our original input file. The PLUMED file for analysis with the driver is:

```
RESTART
MOLINFO MOLTYPE=protein STRUCTURE=egaawaass.pdb
WHOLEMOLECULES ENTITY0=1-111

# CVs go here...
cagyr: GYRATION TYPE=RADIUS ATOMS=5,20,27,37,47,71,81,91,102

PBMETAD ...
ARG=phi2,phi3,phi4,phi5,phi6,phi7,phi8,phi9,psi1,psi2,psi3,psi4
,psi5,psi6,psi7,psi8,dihtrp_cacb,dihtrp_cbcg,aasimpsi,aasimphi,
peplen
SIGMA=0.6,0.6,0.6,0.6,0.6,0.6,0.6,0.6,0.6,0.6,0.6,0.6,0.6,0.6,0
.6,0.6,0.6,0.6,0.3,0.3,0.3
HEIGHT=0.3
PACE=500000000
TEMP=300
BIASFACTOR=8
LABEL=pb
GRID_MIN=-pi,-pi,-pi,-pi,-pi,-pi,-pi,-pi,-pi,-pi,-pi,-pi,-pi,-
pi,-pi,-pi,-pi,-pi,0,0,0
GRID_MAX=pi,pi,pi,pi,pi,pi,pi,pi,pi,pi,pi,pi,pi,pi,pi,pi,pi,pi,
1,1,3.5
... PBMETAD

# We do not need the Metainference directive for post
processing
PRINT ARG=pb.bias FILE=FULLBIAS
PRINT ARG=cagyr FILE=GYR

# Add PRINT directive for all other experimental observables...
PRINT ARG=(jhahn\.j_.*) FILE=JHAHN
```

To perform the analysis, we run the following command:

```
$ plumed driver --plumed plumed-analysis.dat --mf_xtc traj.xtc
```

which will produce the `FULLBIAS` and `GYR` files. These files contain the PBMetaD bias and the radius of gyration for each frame of the trajectory respectively. We also obtain files containing the value of each experimental observable for every frame of the trajectory.

To calculate ensemble averages and free energies, we need to calculate the weight of each frame from the bias, which can be done using the following python code (see Note 2):

```
import numpy as np

KBT = 2.49
bias = np.loadtxt("FULLBIAS")
weights = np.exp(bias[:,1] / KBT)
weights /= weights.sum()
```

Using these unbiasing weights, we can now calculate ensemble averages and probability distributions of any function of the coordinates of the system. We start by calculating ensemble averages for all back-calculated experimental observables. With those, we can obtain root-mean-square deviations (RMSDs) between a particular dataset (such as Hα-HN 3J-couplings) and the experimental reference values (Fig. 7, see Note 14):

We skip loading the first column of the file JHAHN, since it only contains the

```
jhahn = np.loadtxt("JHAHN")[:,1:]
jhahn_mean = (jhahn * weights.reshape(-1, 1)).sum(axis=0)
jhahn_exp = np.array([6.05, 5.95, 6.44, 6.53, 5.93, 6.98, 7.16])
rmsd = np.sqrt(((jhahn_mean - jhahn_exp) ** 2).mean())
```

simulation time, which is not needed for this particular analysis. We then compute the ensemble average using the weights determined above and continue by calculating the RMSD between our ensemble averages and the reference experimental values for a particular dataset. We can use the same principle to compute the RMSDs with respect to the other experimental observables. Looking at the results, we can see a better agreement with the experimental data (Fig. 7).

We will continue by looking at the probability distribution of the radius of gyration

$R_g$. We make use of the weights previously calculated to calculate probability distributions

```
gyr = np.loadtxt("GYR")[:,1]
hist, bins = np.histogram(gyr, bins=50, weights=weights, density=True)
```

The probability densities give us crucial information on the behaviour of the system. In this case, we can see that EGAAWAASS is primarily found in two states (Fig. 8): a fairly compact ($R_g$ ~ 0.5 nm) and a more extended form ($R_g$ ~ 0.8 nm). This feature only emerges when introducing experimental data, as the prior information encoded in the CHARMM22* force field is insufficient to accurately determine these states.

Finally, we also look at the distributions of the Bayesian error $\sigma_{r,i}^B$ for several data points (Fig. 9). In the case of C$\beta$ chemical shifts, we see a fairly large spread for the A6 residue, indicating a relatively weak restraint. This could be due to errors in the parameterization of the CamShift predictor and/or random or systematic errors in the experimental data.

## 5. Notes

1. PBMetaD deposits multiple Gaussians along $n$ one-dimensional CVs as opposed to the one $n$-dimensional Gaussian added in standard metadynamics. As CVs are typically correlated, Gaussians are not simultaneously added to all variables, but only to an "active" CV defined by a discrete switching variable η. After marginalising η, we obtain a conditionally-weighted Gaussian.

2. The weights needed to unbias a PBMetaD simulation can be calculated using the Torrie-Valleau approach *(36)*:

$$w(S_1, S_2) \propto e^{\frac{V_{PB}(S_1, S_2, \bar{t})}{k_B T}}$$

where $V_{PB}(S_1, S_2, \bar{t})$ is the final PBMetaD bias, and $S_1$ and $S_2$ are two CVs.

3. The LINCS constraint parameters used in M&M simulations are typically more conservative than the default of values of 1 iteration and a matrix expansion of the order of 4, because the introduction of experimental data can add additional strains on constrained bonds.

4. The `WHOLEMOLECULES` directive reconstructs inside PLUMED the molecules broken inside the MD code by periodic boundary conditions. The atoms defining the molecules are specified by the `ENTITY0` keyword. Additional molecules can be specified by using multiple `ENTITY` keywords.

5. To run multi-replica simulations, both PLUMED and GROMACS must be compiled with MPI support. The GROMACS executable is typically called `gmx_mpi`, and all simulations should be started with an appropriate launcher, such as `mpiexec` or `mpirun`, and specifying the number of MPI processes to used. The exact name and syntax of this command depends on the system used.

6. The convergence of well-tempered PBMetaD simulations as well as the error in the reconstructed free energies can be assessed by using the block-analysis procedure illustrated in the PLUMED tutorials available at www.plumed.org.

7. For systems with many data points and thus many associated error parameters, one may experience very low MC acceptance rates and thus encounter sampling issues. This problem can be alleviated by performing sampling of the error parameters in groups by using the `MC_CHUNKSIZE` and `MC_STEPS` keywords. For example, with 100 data points, one can perform five MC steps, with 20 data points moved at each step.

8. When setting `DSIGMA`, we need to take care that the step sizes are not too large to cause instabilities and low MC acceptance rate (Fig. 4C) and not too small to result in slow and insufficient sampling.

9. In metainference, $\widetilde{f_i}$ is the (unknown) average of the forward model $f_i$ calculated over an infinite number of replicas, while $p(\widetilde{f}_{r,i}|X, \sigma_{r,i}^{SEM})$ quantifies the difference between the unknown average $\widetilde{f_i}$ and the estimate $f_i$ calculated using a small set of replicas. Therefore, in order to keep $\sigma_{r,i}^{SEM}$ small, we should use as many replicas as possible.

10. The best way to determine the agreement with the experimental data is to calculate the RMSDs of the average of the experimental observables over the entire metainference simulation. However, during the course of the simulation, quantities such as the correlation and RMSDs between the forward model and the experimental value can be used to evaluate the satisfaction of the data-restraints. It is very important to keep in mind that typical values for the correlation might vary between systems and types of data and therefore they can only be used to evaluate the relative as opposed to absolute quality of the fit.

11. In metainference, the sampling of the scaling factor $\lambda$ makes use of an Ornstein-Uhlenbeck process *(55)*. The result is a trajectory of points with limited jumps and sampling a Gaussian distribution:

$$d\lambda_t = \frac{1}{2}(\mu - \lambda_t) + \Delta\lambda dW_t$$

where $d\lambda_t$ is the step taken, $\mu$ is the specified mean of the stationary Gaussian distribution, $\lambda_t$ is the scaling value at time t, $\Delta\lambda$ is the standard deviation of the stationary Gaussian distribution and $dW_t$ denotes the Wiener process (i.e. Brownian motion). Using a process of this form ensures that the step size is not too large and helps to keep the simulation stable.

12. When working with RDCs, one can sometimes observe a negative correlation. In this case one should invert the sign of the scaling factor (in the RDC CV).

13. Specifying the temperature is not required during simulation as it is explicitly passed to PLUMED from the MD engine.

14. The root-mean-square deviation (RMSD) is given by:

$$\text{RMSD} = \sqrt{\frac{1}{N_d} \sum_{i=1}^{N_d} (f_i(\mathbf{X}) - d_i)^2}$$

where $N_d$ is the total number of data points, $f_i(\mathbf{X})$ is the ensemble-average of the forward model for the *i*-th data point and $d_i$ is the reference experimental value.

**Tables**

Table 1: Experimental chemical shifts for the EGAAWAASS peptide (ppm)

| Residue | HN | N | Hα | Cα | C' | Hβ | Cβ |
|---|---|---|---|---|---|---|---|
| E1 | | | 4.103 | 55.83 | 173.15 | 2.152 | 29.99 |
| G2 | 8.780 | 111.42 | 4.034 | 45.12 | 173.46 | | |
| A3 | 8.353 | 124.31 | 4.285 | 52.35 | 177.72 | 1.277 | 19.31 |
| A4 | 8.344 | 123.67 | 4.287 | 52.68 | 177.58 | 1.361 | 19.07 |
| W5 | 8.008 | 119.98 | 4.612 | 57.37 | 175.80 | 3.308 | 29.50 |
| A6 | 7.833 | 126.18 | 4.224 | 52.04 | 176.69 | 1.247 | 19.73 |
| A7 | 8.055 | 123.48 | 4.241 | 52.49 | 177.78 | 1.429 | 19.37 |
| S8 | 8.283 | 115.37 | 4.511 | 58.27 | 173.82 | 3.930 | 64.13 |
| S9 | 8.024 | 122.84 | | 59.91 | 178.50 | | |

Table 2: Experimental RDCs for the EGAAWAASS peptide (Hz)

| Residue | NH | Cα-Hα | Cα C' |
|---|---|---|---|
| E1 | | 12.95 | -0.59 |
| G2 | -5.4 | | -1.55 |
| A3 | -1.26 | 11.5 | -0.67 |
| A4 | -5.22 | 21.42 | -0.94 |
| W5 | -0.91 | -9.37 | -1.49 |
| A6 | 2.33 | 10.01 | -0.55 |
| A7 | -2.88 | 15.01 | -0.3 |
| S8 | -8.37 | 15.73 | -1.44 |
| S9 | -3.78 | | |

Table 3: Experimental 3J-couplings for the EGAAWAASS peptide (Hz)

| Residue | Hα-N | Hα-HN | C-Cγ | N-Cγ |
|---|---|---|---|---|
| E1 | | | | |
| G2 | -0.49 | | | |
| A3 | | 6.05 | | |
| A4 | -0.54 | 5.95 | | |
| W5 | -0.53 | 6.44 | 1.59 | 1.21 |
| A6 | | 6.53 | | |
| A7 | -0.39 | 5.93 | | |
| S8 | -0.39 | 6.98 | | |
| S9 | | 7.16 | | |

# Figures

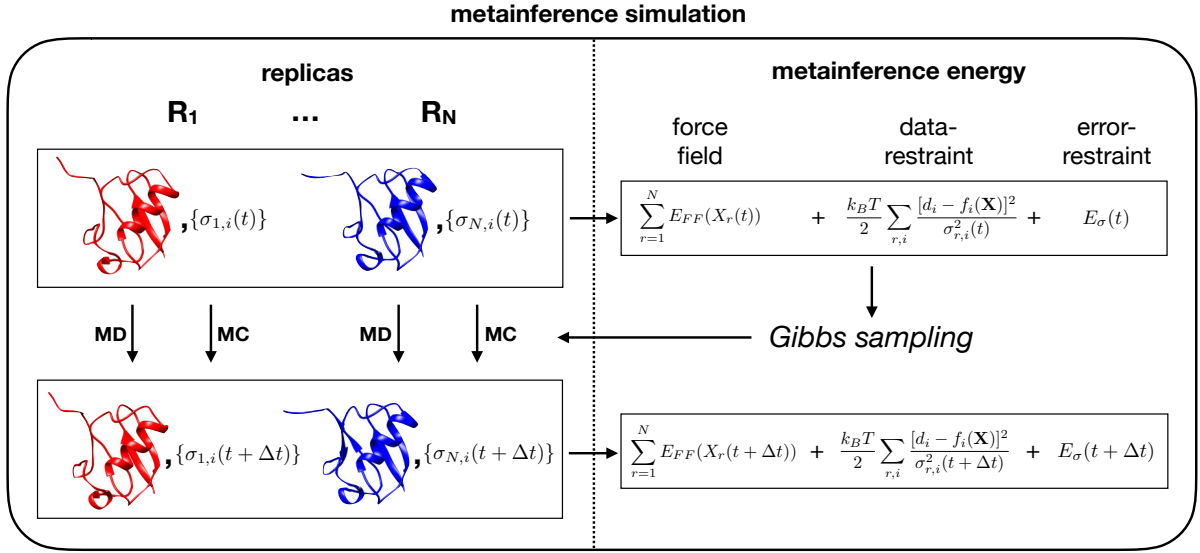

Figure 1. Illustration of the Gibbs sampling mechanism in the multiple-replica MD simulation scheme used in metainference. The metainference energy function $E_{MI}$ is composed by the force field $E_{FF}$, the data-restraints, which enforce the agreement of the forward model averaged across replicas $f_i(\mathbf{X}) = \frac{1}{N}\sum_{r=1}^{N} f_i(X_r)$ with the experimental data, and the error restraint $E_\sigma$ (Eq. 3). The error parameters $\{\sigma_{r,i}\}$ determine the intensity of the data-restraints and are defined as $\sigma_{r,i}^2 = (\sigma_{r,i}^B)^2 + (\sigma_{r,i}^{\text{SEM}})^2$, where $\sigma_{r,i}^B$ is the Bayesian error sampled by MC, and $\sigma_{r,i}^{\text{SEM}}$ the standard error of the mean, which is estimated based on a windowed average (Eq. 4).

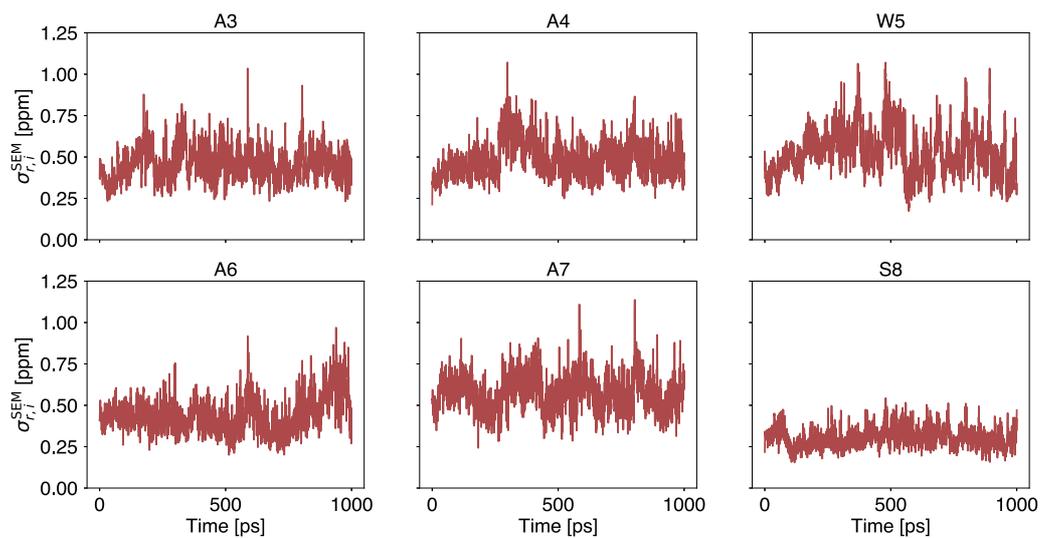

Figure 2. Time series of $\sigma_{r,i}^{\text{SEM}}$ for all the C$\beta$ chemical shifts during the first 1000 ps of simulation. After calculating $\sigma_{r,i}^{\text{SEM}}$ at each time step $t$, the square root of the maximum of this value over the last $m$ steps (200 in our case) is used.

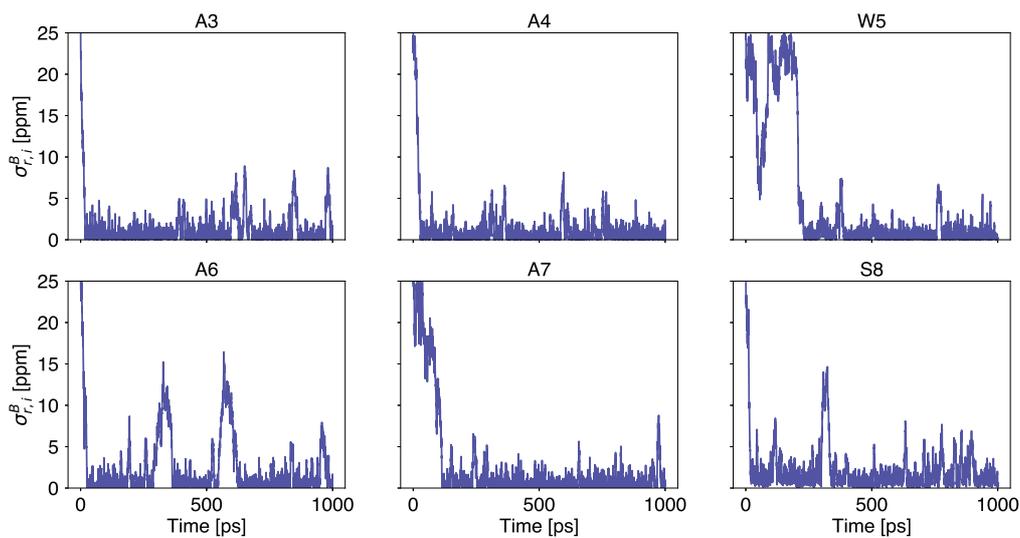

Figure 3. Time series of $\sigma^B_{r,i}$ for all the $C\beta$ chemical shifts during the first 1000 ps of simulation. Their observed decrease in value corresponds to the data-restraint becoming stronger for the corresponding data point. The errors become larger when the structural ensemble is inconsistent with the experimental data.

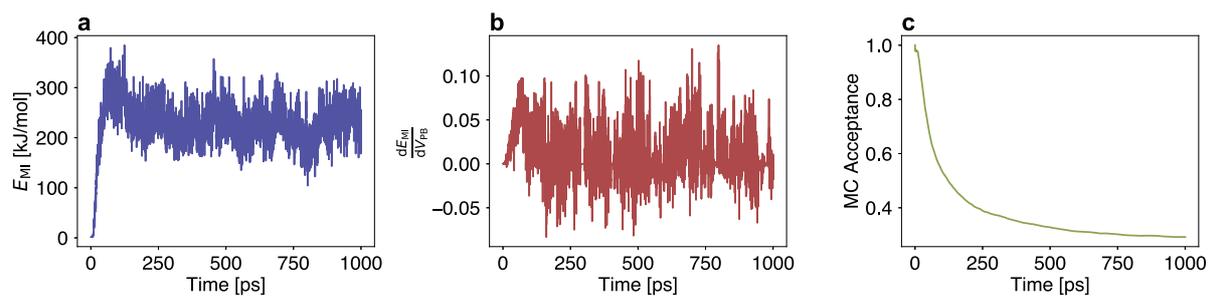

Figure 4. Metainference observables during the first 1000 ps of the M&M simulation with 3J-couplings and chemical shifts. A) Time series of the metainference energy, quantifying the data-restraint intensity. We would typically expect to see a relatively constant value after equilibration. B) The derivative of the metainference energy $E_{MI}$ with respect to the PBMetaD bias $V_{PB}$. C) The average MC step acceptance rate of $\sigma^B_{r,i}$ for all data points.

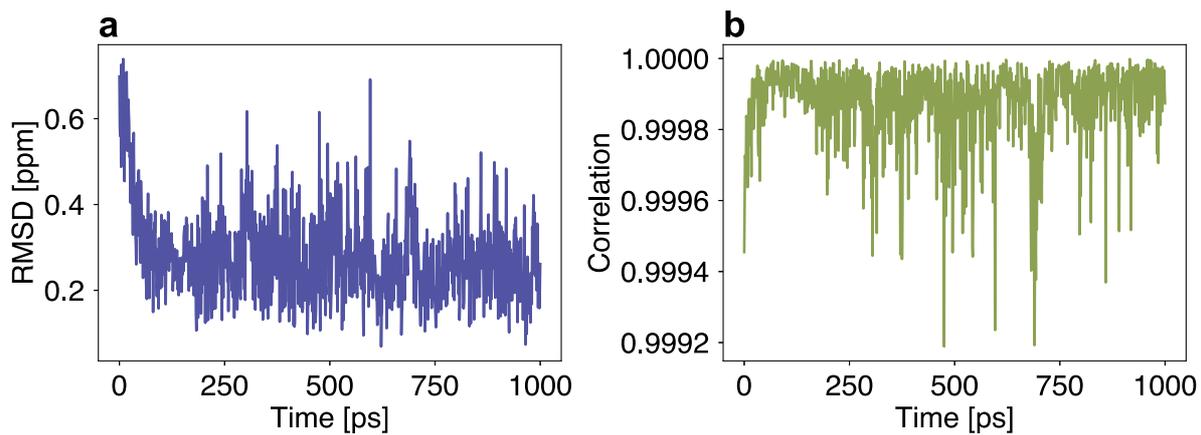

Figure 5. Agreement between calculated and experimental *Cβ* chemical shifts during the first 1000 ps of the simulation. A) RMSD and B) Pearson's correlation coefficient between the back-calculated chemical shifts and the experimental observables.

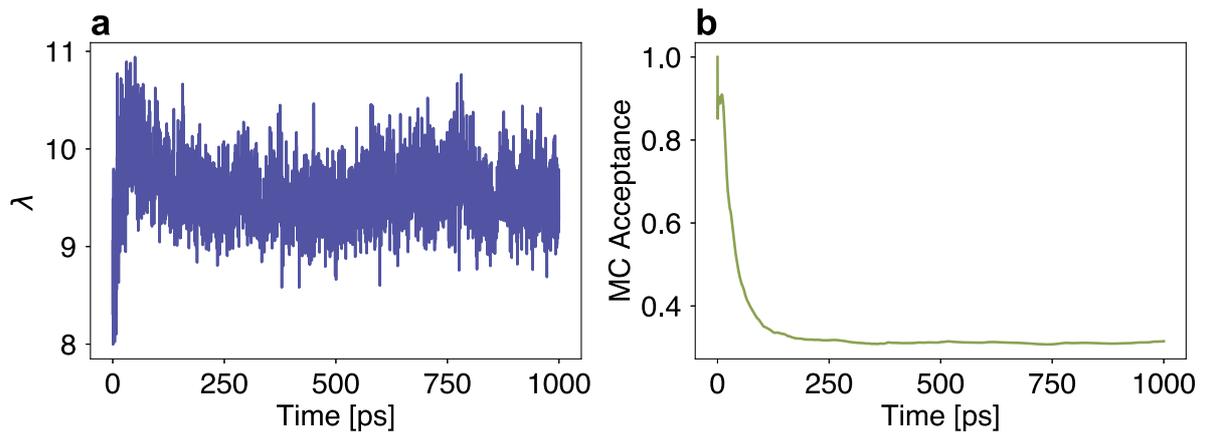

Figure 6. A) The scaling factor $\lambda$ and B) its MC acceptance rate for the NH RDCs. The initial value of $\lambda = 8$ is slightly larger than ideal, as indicated by the mean and the acceptance rate dropping quickly.

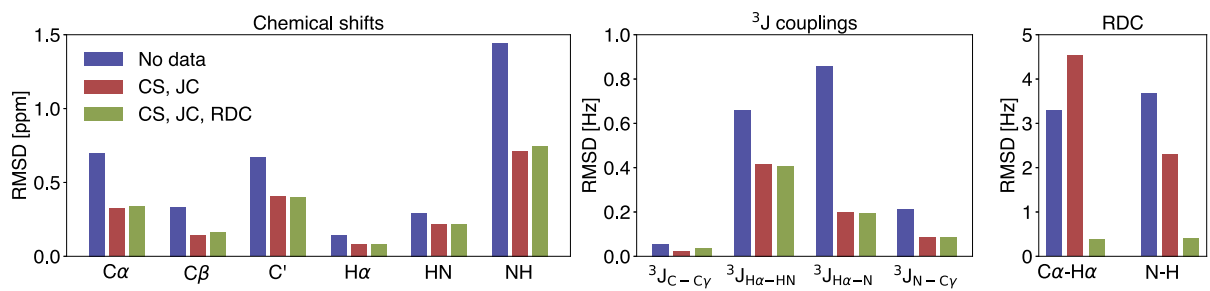

Figure 7. RMSD between the calculated experimental observables from unrestrained, partially restrained (chemical shifts and 3J-couplings) and fully restrained (chemical shifts, 3J-couplings, RDCs) simulations and the experimental measurements. While the agreement of both chemical shifts and 3J-couplings with the experimental data is significantly improved upon the introduction of these data, the RDCs remain largely unaffected. Vice versa, the RDCs have little influence on the quality of the chemical shifts and 3J-couplings.

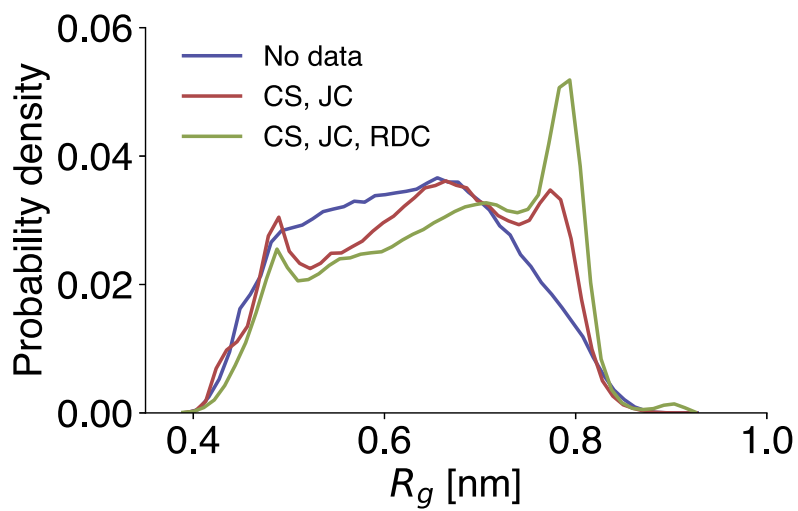

Figure 8. Probability distribution of the radius of gyration $R_g$ for unrestrained, partially restrained (3J-couplings and chemical shifts) and fully restrained (3J-couplings, chemical shifts, and RDCs) simulations. With an increase in experimental data used in metainference, we see the appearance of two distinct peaks from the originally flat distribution.

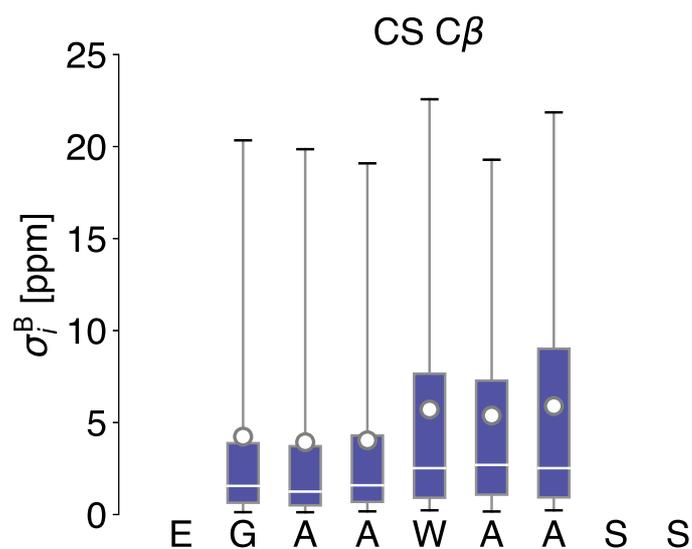

Figure 9. Distribution of $\sigma^B_{r,i}$ for all the C$\beta$ chemical shifts across the metainference ensemble. The mean and median are indicated with a line and a dot, while the box edges and whiskers indicate the mid two quartiles and 5th and 95th percentile respectively. With high-quality data and good sampling, we expect to observe small errors.